\documentclass[conference,pdflatex]{IEEEtran}
\makeatletter
\def\ps@headings{%
\def\@oddhead{\mbox{}\scriptsize\rightmark \hfil \thepage}%
\def\@evenhead{\scriptsize\thepage \hfil \leftmark\mbox{}}%
\def\@oddfoot{}%
\def\@evenfoot{}}
\makeatother
\usepackage[active]{srcltx}
\usepackage{subfigure}
\usepackage{mathptmx}
\usepackage[scaled=-90]{helvet}
\usepackage{courier}
\usepackage{amsmath}
\usepackage{amsfonts,amssymb}
\usepackage{graphicx}

\newtheorem{theorem}{Theorem}

\newtheorem{definition}{Definition}

\newtheorem{fact}{Fact}

\begin{document}

\title{Routing in Outer Space:\\ Improved Security and Energy-Efficiency\\ in
Multi-Hop Wireless Networks}

\author{\IEEEauthorblockN{Alessandro Mei}
\IEEEauthorblockA{Department of Computer Science\\
Sapienza University of Rome, Italy\\
Email: mei@di.uniroma1.it}
\and
\IEEEauthorblockN{Julinda Stefa}
\IEEEauthorblockA{Department of Computer Science\\
Sapienza University of Rome, Italy\\
Email: stefa@di.uniroma1.it}
}

\maketitle
\begin{abstract}
In this paper we consider security-related and energy-efficiency issues in 
multi-hop wireless networks. We start our work from the observation, known in
the literature, that shortest path routing creates congested areas in multi-hop
wireless networks. These areas are critical---they generate both security and
energy efficiency issues. We attack these problems and set out routing in outer
space, a new routing mechanism that transforms any shortest path routing
protocol (or approximated versions of it) into a new protocol that, in case of
uniform traffic, guarantees that every node of the network is responsible for
relaying the same number of messages, on expectation. We can show that a network
that uses routing in outer space does not have congested areas, does not have
the associated security-related issues, does not encourage selfish positioning, and,
in spite of using more energy globally, lives longer of the same network using the
original routing protocol.
\end{abstract}


\section{Introduction}
\label{sec:Introduction}

During the past years the interest in multi-hop wireless networks has been
growing significantly. These types of networks have an important functionality
that is the possibility to use other nodes as relays in order to deliver
messages and data from sources to destinations. This functionality makes
multi-hop wireless networks not only scalable but also usable in various
areas and contexts. One of the most representative and important examples of
multi-hop wireless networks are wireless sensor networks where small devices
equipped with a radio transmitter and a battery are deployed in an
geographic area for monitoring or measuring of some desired property like
temperature, pressure etc.~\cite{akyildiz02survey, habitat02}.

The routing on a wireless sensor network is one of the most interesting and
difficult issues to solve due to the limited resources and capacities of the
nodes. Hence, protocols that use less information possible and need minimal
energy consumption of nodes have become more than valuable in this context. Much
research work has been devoted to finding energy-efficient routing protocols for
this kind of networks. Often, these protocols tend to find an approximation of
the shortest path between the source and the destination of the message.
In~\cite{ksh07}, the authors analyze the impact of shortest path routing in a
large multi-hop wireless network. They show that relay traffic induces congested
areas in the network. If the traffic pattern is uniform, i.\ e.\ every message
has a random source and a random destination uniformly and independently chosen,
and the network area is a disk, then the center of the disk is a congested area,
where the nodes has to relay much more messages that the other nodes of the 
network. We have the same problem if the network area is a square, or a 
rectangle, or any other two-dimensional convex surface. Our experiments show 
that, when using geographic routing~\cite{kk-MOBICOM00} on a network deployed in
a square, 25\% of the messages are relayed by the nodes in a small central
congested region whose area is $3\%$ of the total area of the square. 

Congested areas are bad for a number of important reasons. They raise
security-related issues: If a large number of messages are relayed by the nodes
deployed in a relatively small congested region, then jamming can be a
vicious attack. It is usually expensive to jam a large geographical area,
it is much cheaper and effective to jam a small congested region. In the square,
for example, it is enough to jam 3\% of the network area to stop 25\% of the
messages. Moreover, if an attacker has the goal of getting control over as many
communications as possible, then it is enough to control 3\% of the network
nodes to handle 25\% of the messages.

There are also energy-efficiency issues: Aside from re-transmissions, that are
costly and, in congested areas, often more frequent, the nodes have to relay a
much larger number of messages. Therefore these nodes will die earlier than the
other nodes in the network, exacerbating the problem for the nodes in the 
congested region that are still operational. In the long run, this results in
holes in the network and in a faster, and less graceful, death of the system.
Note that these problems are not solved by trying to balance the load just 
locally, as done by a few protocols in the literature (like
GEAR~\cite{yu01geographical}, for example)---these protocols are useful, they
can be used in any case (in our protocols as well), and are efficient in 
smoothing the energy requirements among neighbors, while they can't do much
against congested areas and they don't help to alleviate the above discussed
security-related issues.

Lastly, there are other concerns in the contexts where the nodes are carried by
individual independent entities. In this paper we do not consider mobility.
However, if the position of the node can be chosen by the node itself in such a
way to maximize its own advantage, and if energy is an issue, then no node would
stay in the center of the square, the highly congested region. If the nodes are
selfish, an uneven distribution of load in the network area leads to an 
irregular distribution of the nodes---there is no point in positioning in the
place where the battery is going to last the shortest. Selfish behavior is a
recent concern in the network community and it is rapidly gaining
importance~\cite{kk-MOBICOM00, sncr-INFOCOM02,wlw-MOBICOM04}. Most of these
contributions show how to devise mechanisms such that selfish nodes can't help
but truthfully execute the protocol. For the best of our knowledge, here we are
raising a new concern, that can be important in mobile networks or whenever the
position of the node can be an independent and selfish choice.

Solving these issues---security, energy-efficiency, and tolerance to (a 
particular case of) selfish behavior---is an important and non-trivial problem,
and, at least partially, our goal. In this paper we attack this problem and set
out \emph{routing in outer space}, a new routing mechanism that transforms any
shortest path routing protocol (or approximated versions of it) into a new 
protocol that, in case of uniform traffic, guarantees that every node of the
network is responsible for relaying the same number of messages, on expectation.
We can show that a network that uses routing in outer space does not have
congested areas, does not have the associated security issues, and, in spite of
using more energy globally, lives longer of the same network using the original
routing protocol---that is, it is more energy-efficient. We support our claims
by showing routing in outer space based on geographic routing, and performing a
large set of experiments.

The rest of the paper is organized as follows: In Section~\ref{sec:related} we
report on the relevant literature in this area; in Section~\ref{sec:solution}
we present the theoretical idea behind our work, we come up with routing in
outer space and prove its mathematical properties; In 
Section~\ref{sec:experiments}, after describing our node and network assumptions
and our simulation environment, we discuss on on the practical issues related
in implementing routing in outer space starting from geographic routing;
lastly, we present an extensive set of experiments, fully supporting our claims.

\section{Related Work}
\label{sec:related}

Routing in multi-hop wireless networks is one of most important, interesting,
and challenging tasks due to network devices limitations and network
dynamics. As a matter of fact this is one of the most studied topics in this
area, and the literature on routing protocols for multi-hop wireless networks
is vast. There have been proposed protocols that maintain routes continuously
(based on distance vector)~\cite{pb94dv,za02dv,sd02dv}, that create routes 
on-demand~\cite{johnson96dynamic,Park97adaptive,pr99dv} or a 
hybrid~\cite{haas97new}. For a good survey and comparison 
see~\cite{broch98comparison,royer99review}. Other examples of routing protocols
for multi-hop wireless networks are those based on link-state like 
OLSR~\cite{jacquet01optimized}, etc.

Geographic routing or position-based routing, where nodes locally decide the next
relay on the basis based on information obtained through some GPS (Global Positioning System)
or other location determination  techniques~\cite{hightower01location}, seems to
be one of the most feasible and studied approach. Examples of reasearch
work on this approach are protocols like GEAR
(Geographical Energy Aware Routing)~\cite{yu01geographical}, GAF (Geographical
Adaptive Fidelity)~\cite{gaf01}. For a good starting survey
see~\cite{saeda04tr}. 

All these protocols try to approximate shortest path between sources and
destinations over the network. In~\cite{pham04performance}, the authors
analytically study the impact of shortest single path routing on a node by
approximating single paths to line segments and characterizing the deviation of
routes from line segments. However, one of the parameters in their model
is not analytically quantified and would need to be estimated via simulations.
In \cite{ganjali04Infocom} the authors introduce an analytical model to
evaluate the imposed load at a node by approximating a shortest path route to a
narrow rectangle, where the load of a node is defined as the number of paths
going through the node. The length of the rectangle is defined as the distance
between source and destination and the width of the rectangle parameterizes
the deviation of paths from the line segments between sources and destinations.
However, since the number of nodes is not parameterized in the analytic model,
the model does not provide an explicit relationship between the load and
the number of nodes. In~\cite{ksh07}, the authors analyze the load for a
homogeneous multi-hop wireless network for the case of straight line routing as
in~\cite{gk00,rajeswaran04capacity,pham04performance,ganjali04Infocom,fdtt06,
ksh07} (where Shortest path routing is approximated to straight line
routing in large multi-hop wireless networks). Assuming uniform traffic, 
in~\cite{ksh07} the authors prove that relays induce so called hotspots or congested
areas in the network. The creation of these types of
congested areas not only affect the overall network throughput by generating
energy-efficiency problems, but also create the serious security problems that
we mentioned before. Of course, geographic routing (which, in dense networks, approximates
the shortest path between source and destination)
also suffers of the same problems. A lot of work has been done regarding to the
energy-efficiency issues, and several approaches that try to solve the problem locally have been proposed
like~\cite{yu01geographical,geraf03zorzi,petrioliGeraf05}.

\section{Routing in Outer Space}
\label{sec:solution}

We model the multi-hop wireless network as a undirected graph $G=(V,E)$, where
$V$ is the set of nodes and $E$ is the set of edges. The nodes are ad-hoc 
deployed on the network area $S$. Formally, it is enough to assume that $S$ is 
a metric space with distance $d_S$ and that every node is a point on $S$.
Given two nodes $u,v\in V$ deployed on~$S$, we will denote the distance between
their positions on the space with $d_S(u,v)$. The nodes have a transmission
range $r$---two nodes $u,v\in V$ are connected by a wireless link $uv\in E$ if
$d_S(u,v)\le r$, that is, their distance is at most $r$. The common practice in
the literature is to take a convex surface as $S$, usually a square, a
rectangle, or a disk, with the usual Euclidean distance. In this paper we assume
that the nodes know their position, either by being equipped with a GPS unit, or
by using one of the many localization protocols~\cite{bulusu-selfconfiguring,
savvidesHanSrivastavaMobiCom01}. And that they know the boundaries of the 
network area $S$; this is possible either by pre-loading this information on the
nodes before deployment, or by using one of the protocols 
in~\cite{fangLocating2006, fekete06, wangBoundaryMobiCom06}.

We started from the observation that shortest path routing on the square, or 
even an approximate version of it, generates congested areas on the center
of the network. We have already discussed that this phenomenon is not desirable.
The same problem is present on the rectangle, on the disk, and
on any two dimensional convex deployment of the network, which is the
common case in practice.
Here, the idea is to relinquish shortest paths so as to get rid of congested
areas, with the goal of improving security, energy efficiency, and tolerance to
selfish behavior of the multi-hop wireless network. As the first step, we have to realize that there do
exist metric spaces that do not present the problem. First, we need a formal
definition of the key property of the metric space we are looking for.

\begin{definition}
\label{def:symmetric}
Consider a multi-hop wireless network deployed on a space $S$. Fix a node $u$
and choose its position on $S$ arbitrarily. Then, deploy the other nodes of the
network uniformly and independently at random. We will say that $S$ is 
\emph{symmetric} if, chosen two nodes $v_1$ and $v_2$ uniformly at random in the
network, the probability that $u$ is on the shortest path from $v_1$ to $v_2$ 
does not depend on its position.
\end{definition}

Clearly, the disk is not a symmetric space as in the above definition.
It has been clearly shown in~\cite{ksh07}---if node $u$ is on the
center of the circle or nearby, the probability that $u$ is traversed by a
message routed along the shortest path from a random source node $v_1$ to a
random destination $v_2$ is larger than that of a node away from the
center of the network area. Clearly, the square has exactly the same problem.
This claim is confirmed by our experiments: $25\%$ of the shortest paths 
traverse a relatively small central disk whose area is $3\%$ of the entire
square.

To solve these problems, our idea is to map the network nodes onto a symmetric
space (the \emph{outer space}) through a mapping that preserves the initial 
network properties (such as distribution, number of nodes, and, with some
limitations, distances between them). The second step is to route messages
through the shortest paths as they are defined on the outer space. When the
outer space and the corresponding mapping are clear from the context, we will
call these paths the \emph{outer space shortest paths}.
Since the outer space is symmetric, we can actually prove that every node in the
network has the same probability of being traversed by an outer space shortest
path. In the following section we will see that, based on this idea, we can 
design practical routing protocols that do not have highly congested areas, 
weaker security, and all the problems we have been discussing here. Now, let's 
make a step back and proceed formally.

Let $S$ be the original space where the network is deployed, and let $T$ be the
outer space, an abstract space we use to describe routes, both metric
spaces with respective distances $d_S$ and $d_T$. We are looking for a mapping
function $\phi : S\mapsto T$ with the following properties:
\begin{enumerate}
 \item
 \label{enum:fair1}
 $\phi$ is an injection;
 \item
 \label{enum:fair2}
 if $x$ is a point taken uniformly at random on~$S$, then $\phi(x)$ is also
 taken uniformly at random on~$T$;
 \item
 \label{enum:fair3}
 for every $r>0$, and every $u,v\in T$, $u\neq v$, if
$d_{T}(\phi(u),\phi(v))\leq r$ then $d_{S}(u,v)\leq r$.
\end{enumerate}
Property~\ref{enum:fair1} guarantees that $\phi^{-1}$ is well-defined,
Property~\ref{enum:fair2} guarantees that a uniform traffic on $S$ is
still a uniform traffic when mapped onto $T$ through $\phi$, and
Property~\ref{enum:fair3} says that paths on $T$ are paths on $S$,
when mapped through $\phi^{-1}$. We'll see later why these properties
are important.

\begin{definition}
\label{def:fair}
A mapping $\phi:S\longrightarrow T$ is \emph{fair} if it enjoys
Properties~\ref{enum:fair1}, \ref{enum:fair2}, and~\ref{enum:fair3}.
\end{definition}

Once such a fair mapping has been fixed, any message from node~$u$ to node~$v$
can be routed following the shortest path $\phi(u),\dotsc,\phi(v)$
between the images of $u$ and $v$ and through the images of the
nodes of the network under $\phi$ on space $T$.
Being $\phi$ a fair mapping, the path $\phi^{-1}(u),\dots,\phi^{-1}(v)$ is a well
defined path on $S$. Indeed, $\phi^{-1}(u)$ is unique for all $u$, since $\phi$ is injective;
and any two consecutive nodes in the shortest path on $T$ are neighbors in $S$ as well,
thanks to Property~\ref{enum:fair3}.
If $T$ is \textit{symmetric} as in Definition~\ref{def:symmetric}, the routing
through $\phi$ would be well distributed over $T$. Hence, this path can be used to route
messages on $S$, giving as a result a homogeneous distribution of the message flow over all the
original network area.
\begin{theorem}
Let $\phi:S\longrightarrow T$ be a 
mapping from source metric space $S$ to target metric space $T$. Assume that 
$\phi$ is fair and $T$ is symmetric. Fixed a node $u\in S$, deployed the other
nodes of the network uniformly at random, and taken a source $v_1\in S$ and a
destination $v_2\in S$ uniformly at random, the probability that the outer
space shortest path $\phi^{-1}(v_1),\dots,\phi^{-1}(v_2)$ traverses $u$ is
independent of the position of $u$ on $S$.
\end{theorem}

The above theorem gives an important hint on how to build a routing protocol 
on a not symmetric network area, in such a way that the message flow is distributed 
homogeneously over all the network. What is needed is to determine a
symmetric space (the outer space) and a fair mapping for it, and then to ``transform''
the shortest paths on the original network area into the corresponding outer space
shortest paths.

We assume that the original network area is a square of side 1. An excellent 
candidate as a symmetric outer space is the torus. A \emph{torus} is a
3-dimensional surface that we can model as $T=[0,t]\times[0,t]$. Let $u_x$ and
$u_y$ be the coordinates of the position of node $u$ on the torus. We can endow
$T$ with the following distance $d_T$:
\begin{equation}
d_T(u,v)=\sqrt{d_x^2+d_y^2},
\end{equation}
where
\begin{align}
d_x&=\min\{\left| u_x - v_x \right|, t-\left| u_x- v_x \right|\},\;\textrm{and}\\
d_y&=\min\{\left| u_y, v_y \right|, t-\left| u_y, v_y \right|\}.
\end{align}
The common way to visualize a torus is to consider a square, and then to fold
it in such a way that the left side is glued together with the right side, and
that the top side is glued together with the bottom side. In the following, we
will picture the torus unfolded, just like a square, as it is commonly done to
easily see this 3-dimensional surface as a 2-dimensional one.
\begin{fact}
A torus surface is \emph{symmetric} as in Definition~\ref{def:symmetric}.
\end{fact}

Clearly, virtually no wireless network in real life is deployed on a torus. Here,
we are using the torus just as an abstract space. We are \emph{not} making any
unreasonable assumption on the nodes of the network being phisically placed
on a torus like area with continuous boundaries, nor are we assuming that the
network area becomes suddenly a torus. Indeed, we assume that
the real network is deployed on the square, where the nodes close to one side
\emph{cannot} communicate with the nodes
close to the opposite side. Crucially, the paths used to deliver the messages
are computed as they are defined through a fair mapping onto the torus, the outer space.
Coming back to our idea, now that the target symmetric outer space has been
chosen, what is left to do is to find a fair mapping $\phi_{ST}$ from the square
to the torus. 

\begin{figure}
\begin{center}
\includegraphics[scale=.45]{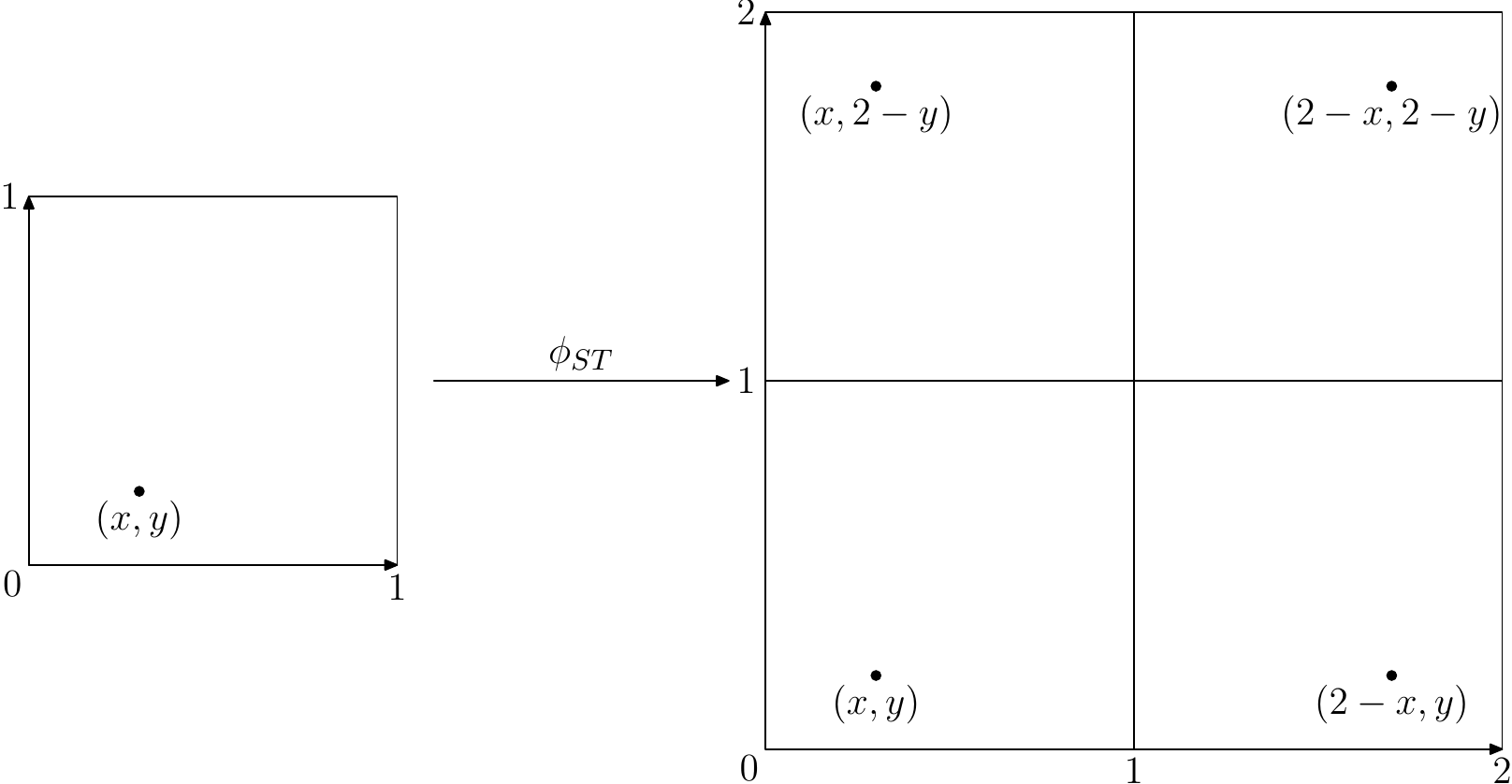}
\end{center}
\caption{Example of transformation of a point from the square to
the torus through the mapping $\phi_{ST}$. Point $(x,y)$ on the square
$S=[0,1]\times [0,1]$ has four possible and equally probable images on the
torus~$T=[0,2]\times [0,2]$. According on $\phi_{ST}$,
only one of the images will actually appear on $T$.}
\label{fig:phi}
\end{figure}

Let $S=[0,1]\times[0,1]$ be a square, and let $T =
[0,2]\times[0,2]$ be a torus. As the mapping $\phi_{ST}$ from $S$ to $T$ we propose
the following:

$ \phi_{ST}((x,y))=(x',y')$ where:
$$x'=
\begin{cases}
x &\textrm{with probability } 1/2\\
2-x &\textrm{with probability } 1/2,
\end{cases}$$
and 
$$y'=
\begin{cases}
y & \textrm{with probability } 1/2\\
2-y &\textrm{with probability } 1/2.
\end{cases}
$$
An example of such a mapping can be seen in Figure~\ref{fig:phi}, where a node
on the square is mapped to one of the four equally probably images on the torus.
\begin{theorem}
$\phi_{ST}$ is a \emph{fair} mapping with probability one.
\end{theorem}
\begin{IEEEproof}
The proof of this claim follows directly from the definition of
the mapping $\phi_{ST}$. The full proof is technical, without
adding much to the understanding of this work. Therefore,
for the sake of brevity, we omit the details.
\end{IEEEproof}

It is interesting to note that it is \emph{not} true that two points that
are neighbors on the square are also neighbors on the torus when
mapped through $\phi_{ST}$. Generally speaking,
it is impossible to build a mapping with both this property
and Property~\ref{enum:fair3}, since the square and the torus are
topologically different.

The outer space shortest path between two nodes may be different
from the corresponding shortest path. Clearly, it can't be shorter by
definition of shortest path on $S$. A natural
question to ask is whether we can bound the stretch, that is, how
much longer may the outer space shortest path be compared with the corresponding
shortest path? Unfortunately, the answer is that the stretch cannot be bounded
by a constant. However, quite surprisingly, we can prove a very good constant
bound in the case when many messages are sent through the network, that is the
common case in practice. Indeed, while in the worst case the stretch can be
high, it is not on average if we assume a uniform traffic. This claim is
formalized in the following theorem, where we show that, on expectation, the
distance of the images under $\phi_{ST}$ of two nodes taken uniformly and
independently at random is at most the double of the original distance.
\begin{theorem}
\label{thm:distances}
If $u,v$ are taken uniformly at random on the square $S=[0,1]\times[0,1]$, and 
$\phi_{ST}(u),\phi_{ST}(v)$ are their respective images under $\phi_{ST}$ on the
torus $T=[0,2]\times[0,2]$, then $$E[d_T(\phi_{ST}(u),\phi_{ST}(v)]\leq
2E[d_S(u,v)].$$
\end{theorem}
\begin{IEEEproof}
Let $u,v\in S$ be two nodes whose position is taken uniformly at random, and let
$E[d_S(u,v)]=\mu$ be the expectation of their distance on $S$. Since $\phi_{ST}$
is fair, also $\phi_{ST}(u)$ and $\phi_{ST}(v)$ are taken uniformly at random in
the torus. Clearly, the distance between $\phi_{ST}(u)$ and $\phi_{ST}(v)$ on 
the torus cannot be larger of the distance of $\phi_{ST}(u)$ and $\phi_{ST}(v)$
on a square $S'=[0,2]\times[0,2]$. Indeed, every path on the torus is also a 
path on the square (the opposite is not true); and the average distance of two
random points in a square of edge two is the double of the average distance of
two random points in a square of edge one. Therefore,
\begin{flalign*}
E[d_T(\phi_{ST}(u),\phi_{ST}(v)]&\le E[d_{S'}(\phi_{ST}(u),\phi_{ST}(v)]\\
&= 2E[d_S(u,v)]\\
&=2\mu.
\end{flalign*}
%
\end{IEEEproof}
In the following, we will see with experiments that the actual average
stretch is even smaller.

Of course, it is always possible to use the outer space shortest path
only when the stretch of that particular path is small, and to use the
classical shortest path when the stretch is high and the outer space
shortest path is going to cost a lot more. However, we do not perform these
kind of optimizations---even though they may reduce the global energy
required by the network to deliver the messages, they also 
unbalance the load among the nodes. In the following, we will
implement our idea in a practical routing protocol derived from
geographical routing, and show its performance by means of
experiments.

\section{Routing in Outer Space in Practice}
\label{sec:experiments}

\begin{figure*}
\centering
\subfigure[Geographic routing between node $u$ and $v$.]{
\centering
\includegraphics[scale=0.5]{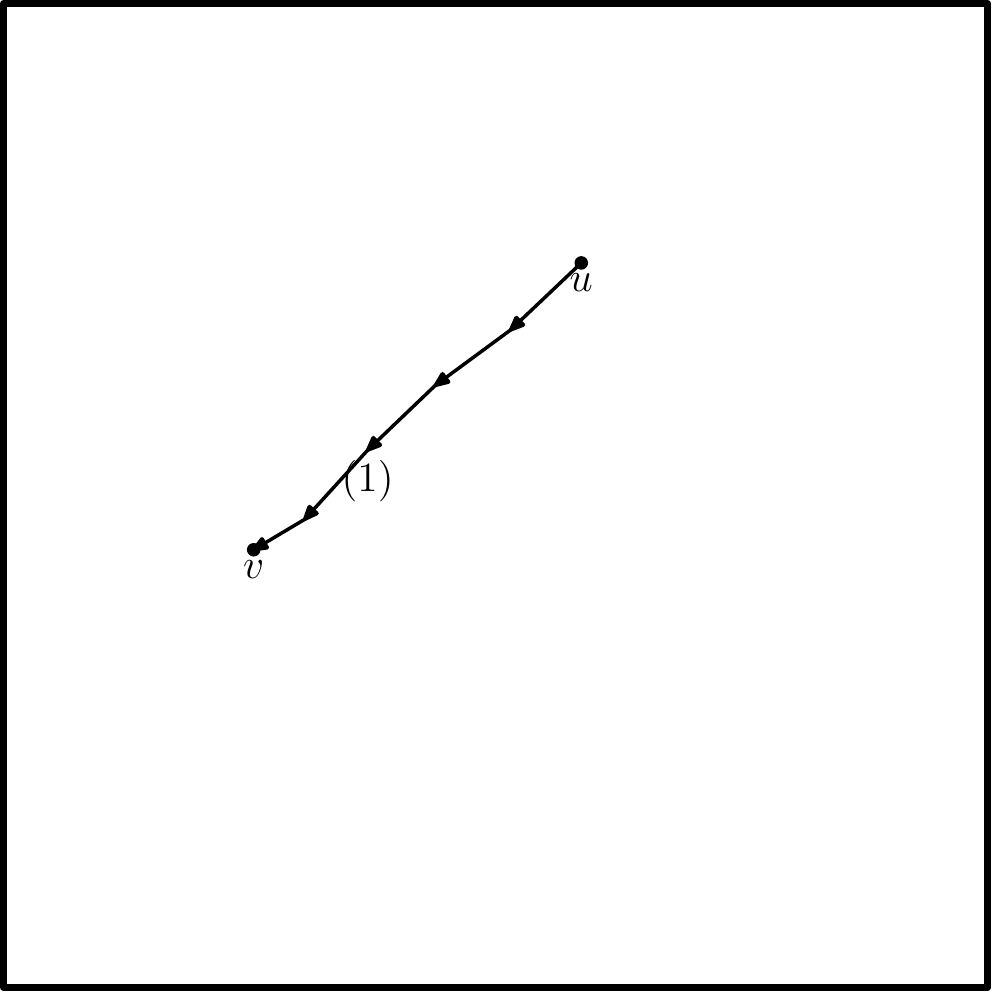}
\label{fig:routeSquare}}
\hspace{1cm}
\subfigure[Four equally probable outer space geographic routings between node $u$ and node $v$.]{
\centering
\includegraphics[scale=0.5]{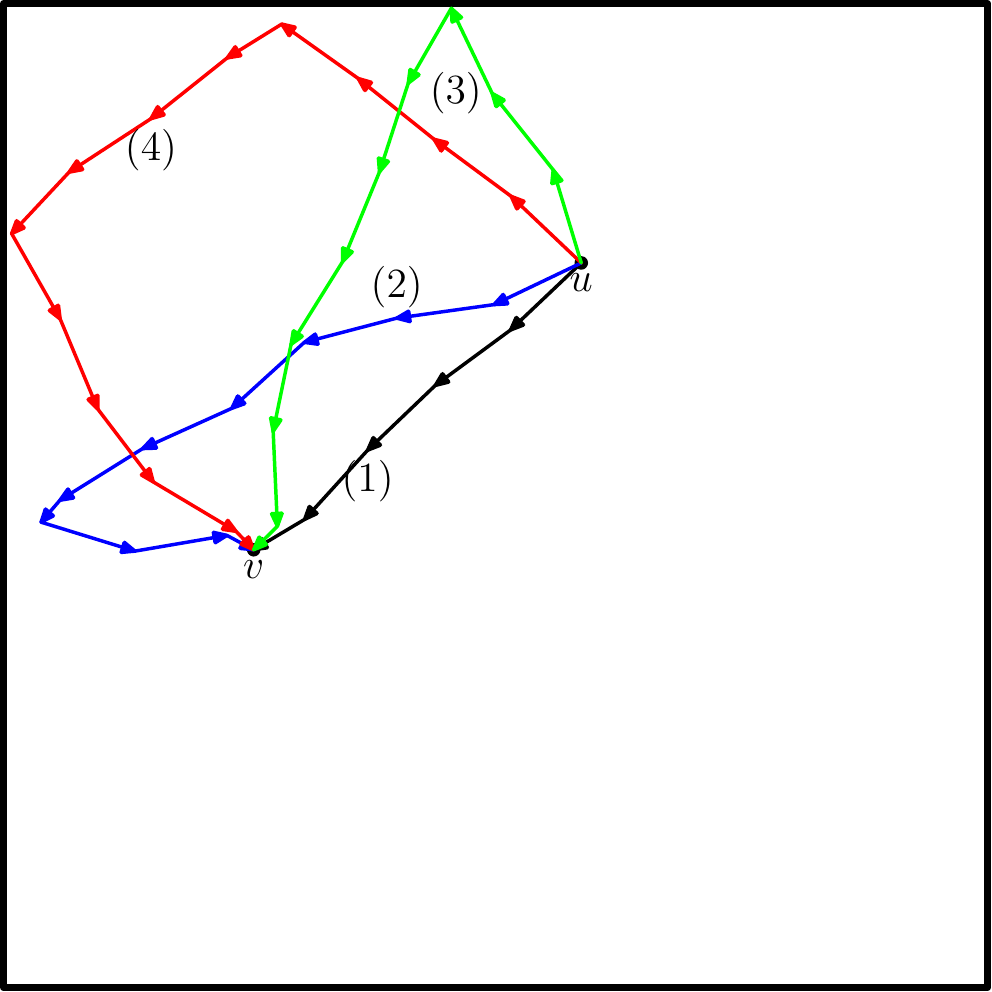}
\label{fig:routeTorus}}
\caption{Transformation of a geographic route on the square into
four possible outer space geographic routes between nodes $u$ and
$v$. As can be seen, depending on which possible $\phi_{ST}(u)$ and
$\phi_{ST}(v)$ images are chosen for $u$ and $v$ on the torus surface, there are
four possible outer space geographic routes on the square between $u$ and
$v$. The network is made of $6,441$ nodes.}
\label{fig:routeChanges}
\end{figure*}

In order to design a practical protocol that makes use of our ideas we start
from geographic routing, a simple protocol that, when the network is dense
enough, can be shown to approximate shortest path routing quite
well~\cite{kk-MOBICOM00}. Here, we define \emph{outer space geographic
routing}, its outer space counterpart.

In geographic routing, the destination of a message is a geographical position 
in the network area---in the square in our case. Every relay node performs a 
very simple protocol: send the message to the node that is closer to 
destination. If such a node does not exist, then the message is delivered. If 
the network is dense, every message is delivered to the node closest to 
destination. It is known that this simple version of geographic routing 
sometimes is not able to deliver the message to the node closest to destination,
and there are plenty of ways to overcome this problem in the literature. 
However, we do not consider these extensions (outer space geographic routing
could as well be based on these more complex and complete versions), since the
increased complexity do not add much to this work.

Outer space geographic routing works quite as simply. Every relay node looks
at the destination $x$ of the message, and forwards it to the node~$u$ that
minimizes $d_T(\phi_{ST}(x),\phi_{ST}(u))$. Just like geographic routing, implemented on
the outer space. 

Take, as an example, a message from node $u$ destined to a geographic position
close to node~$v$. According to the definition of $\phi_{ST}$, each node on the
square~$S$ has four possible and equally probable images on the torus~$T$. This
implies that for each pair $u$, $v$ of nodes on $S$ there are four possible and
equally probable pairs of images $\phi_{ST}(u)$, $\phi_{ST}(v)$ on 
$T$\footnote{Actually, there are 16 possible and equiprobable such couples up to
isomorphism, which fall into 4 different classes of symmetry.}. This yields four
possible and different outer space geographic routes between the images $u$ and
$v$ under $\phi_{ST}$. Hence, between any two nodes on the square there is one
out of four different and equally probable outer space routes. To see an example
of the four routes, see  Figure~\ref{fig:routeChanges}.

To implement such a routing, it is enough that the nodes know their position in the
square. Then, computing $\phi_{ST}$ for itself and the neighbors is trivial and fast.
Note that it is not really important that the nodes agree on which of the four 
possible images is actually chosen for any particular node (except for the 
destination, but the problem can easily be fixed). However, to get this 
agreement it is enough that every node uses the same pseudo-random number
generator, seeded with the id of the node being mapped.

\subsection{Node and Network Properties, Assumptions, and Simulation Environment}
\label{experiments:intro}

We model our network node as a sensor. A typical example can be the Mica2DOT
node (outdoor range $150\textrm{m}$, 3V coin cell battery). These nodes have
been widely used in sensor network academic research and real testbeds.  For our
experiments, we have considered networks with up to 10,000 nodes, distributed on
a square of side $1,500\textrm{m}$. In the following, we will assume for the 
sake of simplicity that the side of the square is 1, and that the node 
transmission range is 0.1. The nodes are placed according to a Poisson 
distribution with density $\rho$, chosen in such a way that every node has 
30--40 neighbors on average. This is a reasonable and commonly used density for
this kind of networks.

We inject a \emph{uniform traffic} in the network---every message has a random
source and a random destination uniformly and independently chosen.
This type of traffic distribution is highly used in the simulations, for example
when the goal is to study network capacity limits, optimal routing, and
security properties~\cite{gk00, zv05, HWKZL05}. We assume that the nodes know 
their position on the network area. Therefore, they need to know both their 
absolute position, and their position within the square. The nodes can get the 
absolute position either in hardware, by using a GPS (Global Positioning 
System),  or in software. There exist several techniques for location sensing 
like those based on proximity or triangulation using different types of signals
like radio, infrared acoustic, etc. Based on these techniques, several location
systems have been proposed in the literature like infrastructure-based 
localization systems~\cite{ward97new, priyantha00cricket} and ad-hoc
localization systems~\cite{bulusu-selfconfiguring, 
savvidesHanSrivastavaMobiCom01}. In~\cite{hightower01location} you can find a 
survey on these systems. Once the absolute position is known, we can get the
nodes to know their relative position within the square by pre-loading the 
information on the deployment area, or by using one of the several techniques 
for boundary detection based on geometry methods, statistical methods, and
topological methods (see~\cite{fangLocating2006, fekete06,
wangBoundaryMobiCom06}).

In the next two sections we present the results of the experiments we have
performed, comparing our routing scheme with geographic routing over the same 
networks and with the same set of messages to route. For the experiments we have
used our own event-based simulator. The assumptions and network properties 
listed above have been exactly reflected in the behavior of the simulator.

\subsection{Security-Related Experiments}
\label{experiments:security}

\begin{figure*}
\centering
\subfigure[Geographic routing.]{
\centering
\includegraphics[scale=0.5]{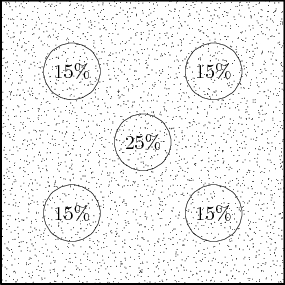}
\label{fig:securitySquare}}
\hspace{1cm}
\subfigure[Outer space geographic routing.]{
\centering
\includegraphics[scale=0.5]{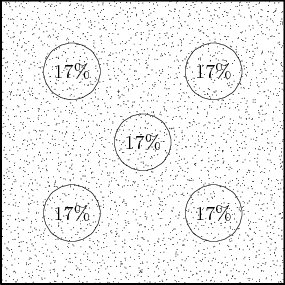}
\label{fig:securityTorus}}
\caption{The average fraction of the messages
whose routing path traverses the selected sub-areas of a network of $1,336$
nodes, in the case of geographic routing and in the case of outer space 
geographic routing.}
\label{fig:securityResult}
\end{figure*}
In these experiments, we measure the number of messages whose routing path
traverses five sub-areas of the same size in the network area.
Every sub-area is a circle of radius $0.1$ (incidentally, the same of the
transmission radius of a network node), that corresponds to an area of $3.14\%$ of the
whole network surface. The sub-areas are centered in some ``crucial'' points of the network
area: The center  and the middle-half-diagonals points. The center of the network is
known to be the most congested area. We want to test whether the middle-half-diagonal
centered areas handle a significantly smaller number of messages.
More specifically we consider the sub-areas centered in the points of coordinates
$(0.5,0.5)$, $(0.25,0.25)$, $(0.25,0.75)$, $(0.75,0.25)$, $(0.75,0.75)$, assuming
a square of side one. Our experiments are done on networks with different number of
nodes (from 1,000 to 10,000). For each  network we have launched both geographic
routing and outer space geographic routing on message sets of different
cardinality (from $50,000$ to $1,000,000$ of messages, generated as an instance
of uniform traffic). In Figure~\ref{fig:securityResult} we present the average 
of the results obtained with a network of $1,336$ nodes generated by a Poisson 
process, but we stress out that exactly the same results are obtained for 
networks with up to $10,000$ nodes. As it can be seen, the experiments fully
support the findings in~\cite{ksh07}. Geographic routing (see 
Figure~\ref{fig:securitySquare}) concentrates a relevant fraction of the 
messages on a small central area of the network, while the other sub-areas 
handle on average little more than the half. We have already discussed why this
is dangerous, and important to avoid.

Figure~\ref{fig:securityTorus} shows the result with the same set of messages
and the same network deployment, this time using outer space geographic
routing. The message load in the central sub-area is $30\%$ lower compared
with the load of the same sub-area in the case of the geographic routing.
Outer space geographic routing seems to transform the network
area in a symmetric surface, making sure that the number of message handled
by all the sub-areas remains reasonably low, $17\%$, and equally distributed.
As a result, the load among network nodes is equally balanced and there are
no ``over-loaded'' areas. This network is intuitively stronger than the same
network using geographic routing, there are no areas that are clearly more
rewarding as objective of a malicious attack, and no network areas have
more ``responsabilities'' than others.

Furthermore, Figure~\ref{fig:securitySquare} clearly shows that, with 
geographic routing, it is not a good strategy to stay in the center of the 
network if you want to save your battery. If the nodes are selfish, it is a
much better strategy to position in one of the sub-central areas, for example,
where the battery is going to last $66\%$ longer. Even better if you move
towards the side of the square. Conversely, when using outer space geographic
routing, there is no advantage in choosing one position or the other, which is
exactly our goal to guarantee an even distribution of the nodes, although part
of them are selfish.

\subsection{How to Live Longer by Consuming More Energy}
\label{experiments:lifetime}

\begin{figure}
\begin{center}
\includegraphics[scale=.4]{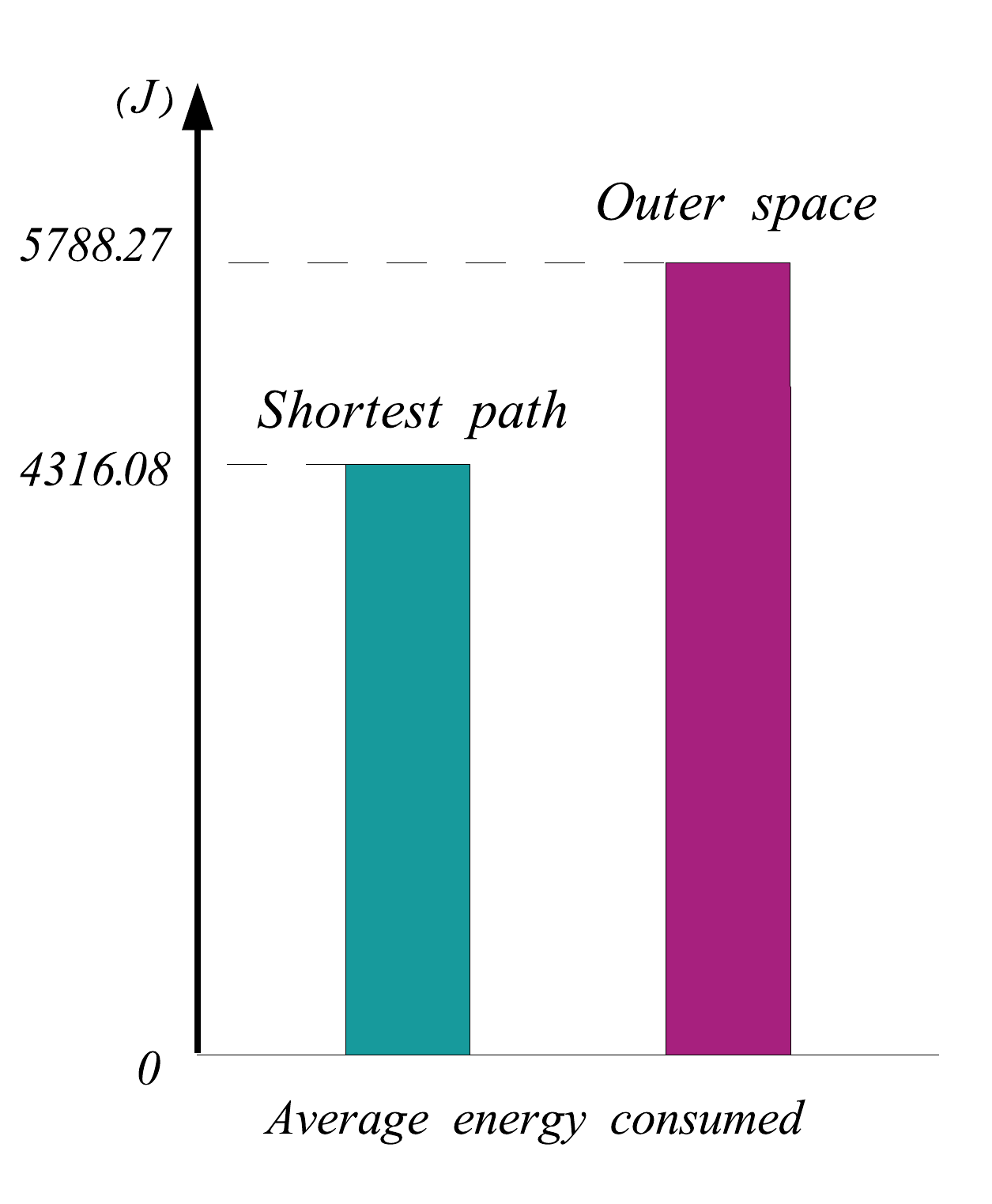}
\end{center}
\caption{Global energy consumption of the network nodes after
running geographic and outer space geographic routing, respectively, on
sets of $50,000$ messages each.The network considered is made of
$1,625$ nodes.}
\label{fig:energy}
\end{figure}

Our main motivation is related to security. However, it is always important
to understand what is the energy overhead of getting rid of congested areas.
Indeed, what Theorem~\ref{thm:distances} says in a word is that the paths using
outer space geographic routing are on average (at most) twice as long as the
paths using geographic routing. This should have an immediate consequence on 
energy consumption: Messages routed with outer space geographic routing should 
make network nodes consume more energy, up to twice as much. And it actually is
 so. What it turns out with our experiments is that the overall energy 
consumption is about $1.4$ times larger with outer space geographic routing, see
Figure~\ref{fig:energy}. Like before, the figure shows the result with a network
of $1,625$ nodes, but we have done more experiments with different sizes, up to 
$10,000$ nodes, and the result does not change.

\begin{figure}
\begin{center}
\includegraphics[scale=.4]{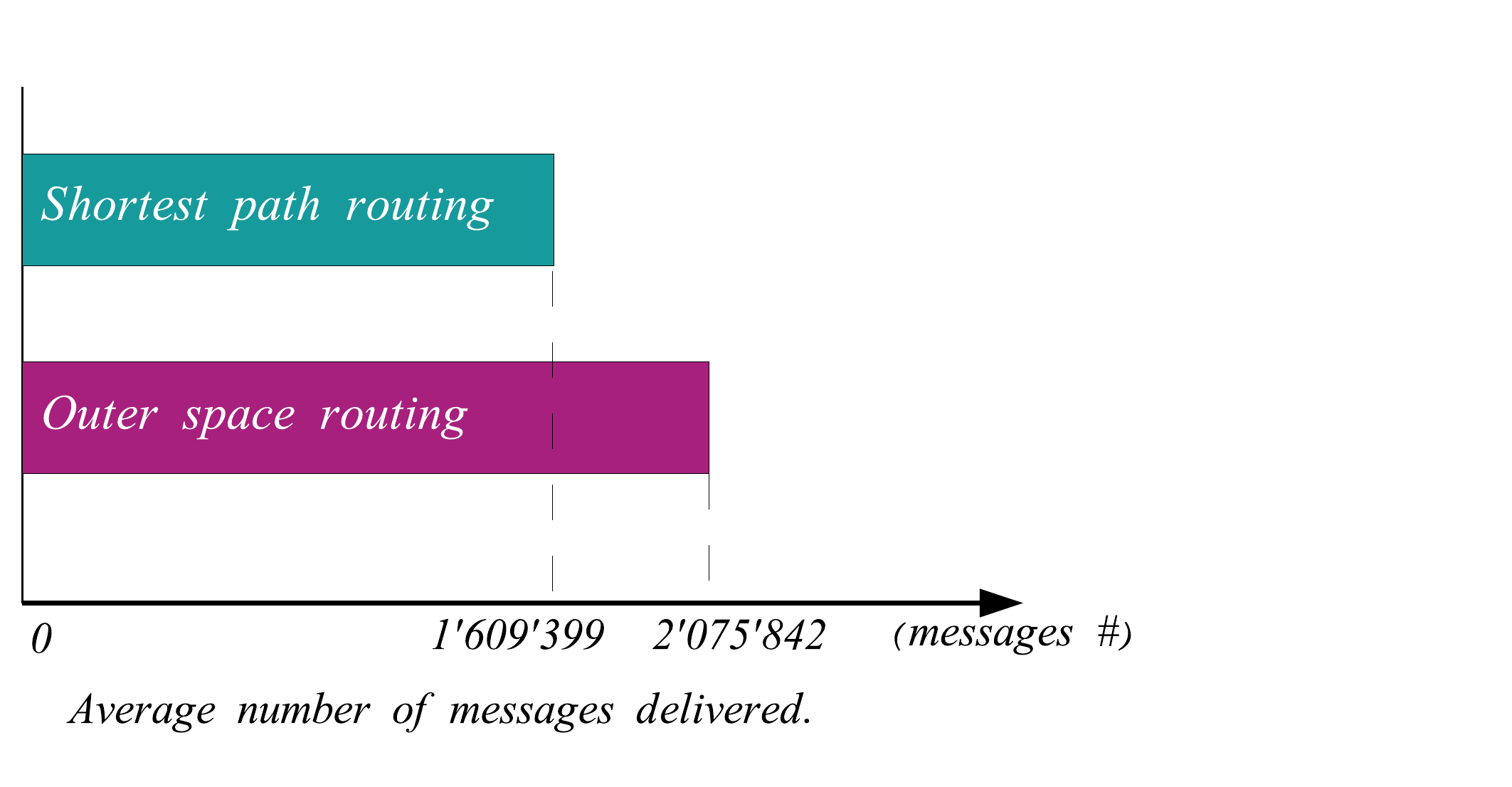}
\end{center}
\caption{Time to first node death. The time is measured as the number of
messages delivered to destination before the death of the first node. The
network consists of $1,625$ nodes.}
\label{fig:firstNodeDeath}
\end{figure}

Usually, when a wireless network consumes more energy, its life is shorter.
However, it is not always the case. Sometimes it is better to consume more
energy, if this is done more equally in the network. This is exactly what
happens with outer space geographic routing. We consider two measure
of network longevity: time to first node death, time to loss of efficiency in
delivery messages. These measures are well-known and used in the
literature~\cite{bhardwajChandraksanInfocom02, bloughSantiMobicom02,
zhangHouMOBIHOC04}.
We have made two sets of experiments, each using one of the above
way to measure the longevity of the network. In each of the experiments
we count the number of messages that are successfully delivered before
network ``death'', where network death is defined according to the above
two measures.
The first set of experiments is done according to the first measure. We have
generated the network, the uniform traffic, and injected the traffic into two 
copies of the same network, one using geographic routing and one using outer
space geographic routing. This have been iterated several times with networks
of different sizes. The result is shown in Figure~\ref{fig:firstNodeDeath},
where we show the number of messages delivered on average by a network of 
$1,625$ nodes (the result does not change by considering network of different 
size), using the two routing protocols under evaluation.
As you can see, the network lifetime of outer space geographic routing
is $22.57\%$ longer, on average, than simple geographic routing. 
As a matter of fact, the number of messages successfully delivered by the
network until the very first node death is much greater with outer space
routing.
\begin{figure}
\begin{center}
\includegraphics[scale=.39]{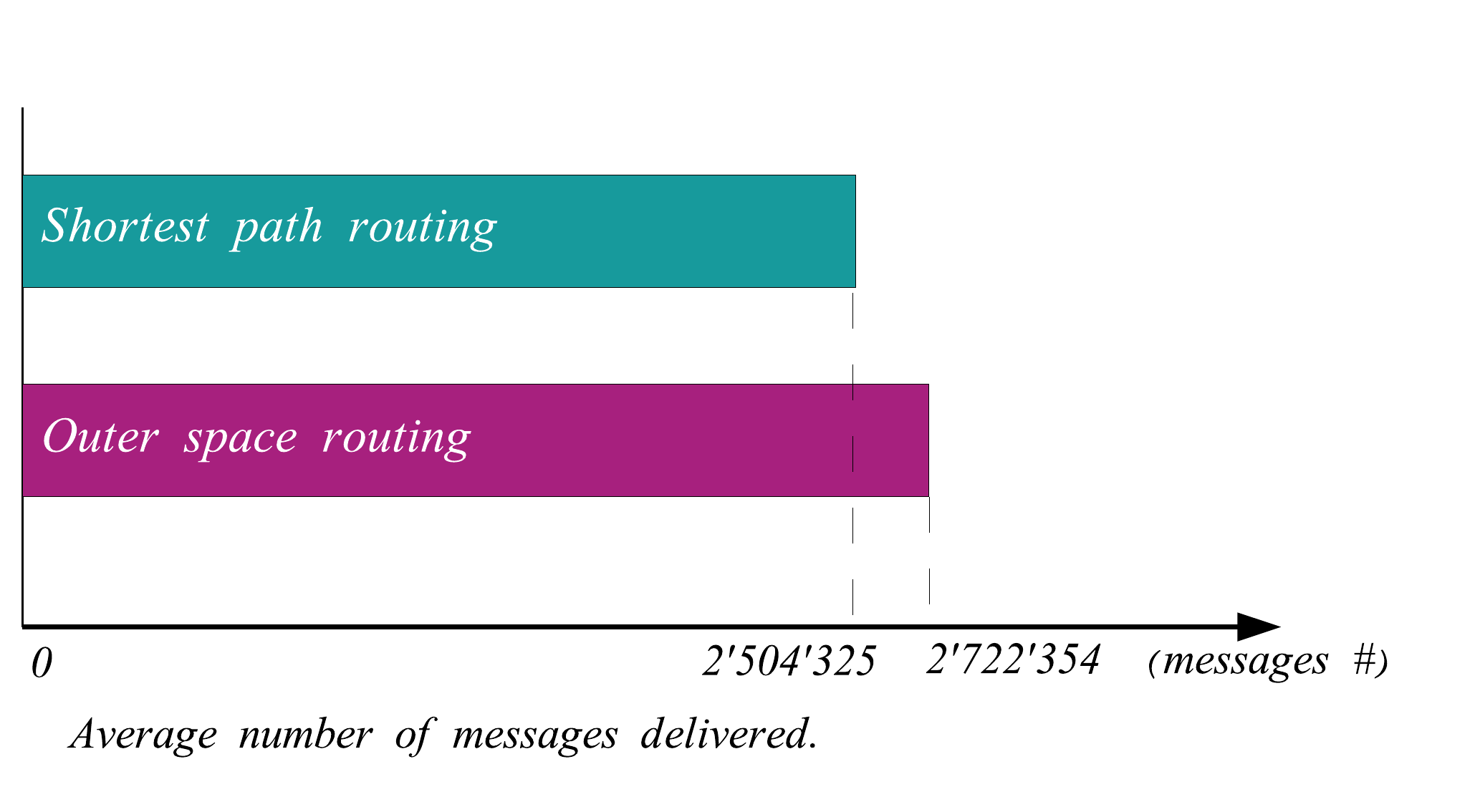}
\end{center}
\caption{Time to delivered percentage. The time is measured as the number of
messages delivered to destination before the delivery success ratio falls
under $95\%$. The network considered is made of $1,625$ nodes.}
\label{fig:deliveredPercentage}
\end{figure}
Since security usually comes at a price, this is somewhat surprising. Routing in
outer space seems to deliver a more secure routing and a more energy-efficient network,
simultaneously.

Figure~\ref{fig:deliveredPercentage} shows the result we get when considering
the second definition of network lifetime. In this case, we consider the network
dead when it is not efficient any more in delivering messages. Note that geographic
routing (and similarly its outer space version) has the problem of ``dead ends'',
places where the message cannot proceed because there is no node closer
to destination, while the destination is still far. There are a number of solutions
to this problem, and there do exists more sophisticated versions of geographic
routing that know how to detect this situation and deliver the message
whenever there is a path between source and destination. However, this
mechanisms are usually costly. When the network is not able any longer to
deliver messages with simple geographic routing, that means that there
has been enough deterioration to create many dead ends in the network
itself. We use this as a measure of the quality of its structure. In this set
of experiments we count the number of messages that reach destination until the
percentage of delivery falls under some threshold (in our case $95\%$).
As can be seen in the figure, even in this case outer space geographic routing
prolongs the life of the network to a value that is on average $11.14\%$ larger
than
the one achieved with geographic routing.
%

\section{Conclusions}

Uniform traffic injected into multi-hop wireless networks generates congested
areas. These areas carry a number of non-trivial issues about security,
energy-efficiency, and tolerance to (a particular case of) selfish behavior.
In this paper we describe routing in outer space, a mechanism to
transform shortest path routing protocols into new protocols that do not
have the above mentioned problems.

Routing in outer space guarantees that every node of the network is responsible for
relaying the same number of messages, on expectation. We can show that a network
that uses routing in outer space does not have congested areas, does not have
the associated security-related issues, does not encourage selfish positioning,
and, in spite of using more energy globally, lives longer of the same network using
the original routing protocol.

\bibliographystyle{ieeetr}
\bibliography{biblio}

\begin{thebibliography}{10}

\bibitem{akyildiz02survey}
I.~Akyildiz, W.~Su, Y.~Sankarasubramaniam, and E.~Cayirci, ``A survey on sensor
  networks,'' {\em IEEE Communcations Magazine}, vol.~40, pp.~102--114, August
  2002.

\bibitem{habitat02}
A.~Mainwaring, D.~Culler, J.~Polastre, R.~Szewczyk, and J.~Anderson, ``Wireless
  sensor networks for habitat monitoring,'' in {\em WSNA '02: Proceedings of
  the 1st ACM international workshop on Wireless sensor networks and
  applications}, (New York, NY, USA), pp.~88--97, ACM Press, 2002.

\bibitem{ksh07}
S.~Kwon and N.~B. Shroff, ``Paradox of shortest path routing for large
  multi-hop wireless networks,'' in {\em IEEE INFOCOM'07 Anchorage}, May 2007.

\bibitem{kk-MOBICOM00}
B.~Karp and H.~T. Kung, ``Gpsr: Greedy perimeter stateless routing for wireless
  networks,'' in {\em Proceedings of the 6th Annual ACM/IEEE International
  Conference on Mobile Computing and Networking (MobiCom '00)}, (Boston, MA,
  USA), August 2000.

\bibitem{yu01geographical}
Y.~Yu, R.~Govindan, and D.~Estrin, ``Geographical and energy aware routing: A
  recursive data dissemination protocol for wireless sensor networks,'' Tech.
  Rep. UCLA/CSD-TR-01-0023, UCLA Computer Science Department, May 2001.

\bibitem{sncr-INFOCOM02}
V.~Srinivasan, P.~Neggehalli, C.~F. Chiasserini, and R.~R. Rao, ``Cooperation
  in wireless ad hoc wireless networks,'' in {\em Proceedings of the
  Twenty-Second Annual Joint Conference of the IEEE Computer and Communications
  Societies (INFOCOM 2003)}, 2003.

\bibitem{wlw-MOBICOM04}
W.~Wang, X.~Li, and Y.~Wang, ``Truthful multicast routing in selfish wireless
  networks,'' in {\em Proceedings of the 10th Annual ACM/IEEE International
  Conference on Mobile Computing and Networking (MobiCom '04)}, (Philadelphia,
  PA, USA), September 2004.

\bibitem{pb94dv}
C.~E. Perkins and P.~Bhagwat, ``Highly dynamic destination-sequenced
  distance-vector routing (dsdv) for mobile computers,'' in {\em SIGCOMM '94:
  Proceedings of the conference on Communications architectures, protocols and
  applications}, (New York, NY, USA), pp.~234--244, ACM Press, 1994.

\bibitem{za02dv}
M.~G. Zapata and N.~Asokan, ``Securing ad hoc routing protocols,'' in {\em
  Proceedings of the ACM Workshop on Wireles Security (WiSe)}, 2002.

\bibitem{sd02dv}
K.~Sanzgiri, B.~Dahill, B.~Levine, C.~Shields, and E.~Belding-Royer, ``A secure
  routing protocol for ad hoc networks,'' in {\em Proceedings of the
  International Conference on Network Protocols (ICNP)}, 2002.

\bibitem{johnson96dynamic}
D.~B. Johnson and D.~A. Maltz, ``Dynamic source routing in ad hoc wireless
  networks,'' in {\em Mobile Computing} (Imielinski and Korth, eds.), vol.~353,
  Kluwer Academic Publishers, 1996.

\bibitem{Park97adaptive}
V.~D. Park and M.~S. Corson, ``A highly adaptive distributed routing algorithm
  for mobile wireless networks,'' in {\em INFOCOM '97: Proceedings of the
  INFOCOM '97. Sixteenth Annual Joint Conference of the IEEE Computer and
  Communications Societies. Driving the Information Revolution}, (Washington,
  DC, USA), p.~1405, IEEE Computer Society, 1997.

\bibitem{pr99dv}
C.~Perkins and E.~Royer, ``Ad hoc on-demand distance vector routing,'' in {\em
  Proceedings of the IEEE Workshop on Mobile Computing Systems and
  Applications}, pp.~90--100, February 1999.

\bibitem{haas97new}
Z.~Haas, ``A new routing protocol for the reconfigurable wireless networks,''
  in {\em Proc. of the IEEE Int. Conf. on Universal Personal Communications},
  October 1997.

\bibitem{broch98comparison}
J.~Broch, D.~A. Maltz, D.~B. Johnson, Y.-C. Hu, and J.~Jetcheva, ``A
  performance comparison of multi-hop wireless ad hoc network routing
  protocols,'' in {\em MobiCom '98: Proceedings of the 4th annual ACM/IEEE
  international conference on Mobile computing and networking}, (New York, NY,
  USA), pp.~85--97, ACM Press, 1998.

\bibitem{royer99review}
E.~Royer and C.~Toh, ``A review of current routing protocols for ad-hoc mobile
  wireless networks,'' April 1999.

\bibitem{jacquet01optimized}
P.~Jacquet, P.~M{\"u}hlethaler, T.~Clausen, A.~Laouiti, A.~Qayyum, and
  L.~Viennot, ``Optimized link state routing protocol for ad hoc networks,'' in
  {\em Proceedings of the 5th IEEE Multi Topic Conference (INMIC 2001)}, 2001.

\bibitem{hightower01location}
J.~Hightower and G.~Borriello, ``Location systems for ubiquitous computing,''
  {\em IEEE Computer}, vol.~34, pp.~57--66, August 2001.

\bibitem{gaf01}
Y.~Xu, J.~Heidemann, and D.~Estrin, ``Geography-informed energy conservation
  for ad hoc routing,'' in {\em Proceedings of the 7th Annual ACM/IEEE
  International Conference on Mobile Computing and Networking (MobiCom '01)},
  (Rome, Italy), July 2001.

\bibitem{saeda04tr}
K.~Seada and A.~Helmy, ``Geographic protocols in sensor networks,'' tech. rep.,
  USC, July 2004.

\bibitem{pham04performance}
P.~P. Pham and S.~Perreau, ``Increasing the network performance using
  multi-path routing mechanism with load balance,'' {\em Ad Hoc Networks},
  vol.~2, pp.~433--459, October 2004.

\bibitem{ganjali04Infocom}
Y.~Ganjali and A.~Keshavarzian, ``Load balancing in ad hoc networks:
  single-path routing vs. multi-path routing,'' in {\em Proceedings of the
  Twenty-third Annual Joint Conference of the IEEE Computer and Communications
  Societies (INFOCOM 2004)}, vol.~2, pp.~1120--1125 vol.2, 2004.

\bibitem{gk00}
P.~Gupta and P.~R. Kumar, ``The capacity of wireless networks,'' {\em IEEE
  Transactions On Information Theory}, vol.~46, March 2000.

\bibitem{rajeswaran04capacity}
A.~Rajeswaran and R.~Negi, ``Capacity of power constrained ad-hoc networks,''
  in {\em Proceedings INFOCOM, 2004.}, vol.~1, pp.~7--11, March 2004.

\bibitem{fdtt06}
M.~Franceschetti, O.~Dousse, D.~Tse, and P.~Thiran, ``On the throughput
  capacity of random wireless networks,'' {\em IEEE Transactions on Information
  Theory}, vol.~52, june 2006.

\bibitem{geraf03zorzi}
M.~Zorzi and R.~R. Rao, ``Geographic random forwarding (geraf) for ad hoc and
  sensor networks: Energy and latency performance,'' {\em IEEE Transactions on
  Mobile Computing}, vol.~2, no.~4, pp.~349--365, 2003.

\bibitem{petrioliGeraf05}
P.~Casari, A.~Marcucci, M.~Nati, C.~Petrioli, and M.~Zorzi, ``A detailed
  simulation study of geographic random forwarding (geraf) in wireless sensor
  networks,'' in {\em Military Communications Conference, 2005. MILCOM
  2005.IEEE}, vol.~1, pp.~59--68, October 2005.

\bibitem{bulusu-selfconfiguring}
N.~Bulusu, J.~Heidemann, D.~Estrin, and T.~Tran, ``Self--configuring
  localization systems: Design and experimental evaluation,'' {\em Trans. on
  Embedded Computing Sys.}, vol.~3, no.~1, pp.~24--60, 2004.

\bibitem{savvidesHanSrivastavaMobiCom01}
A.~Savvides, C.-C. Han, and M.~B. Srivastava, ``Dynamic fine-grain localization
  in ad-hoc networks of sensors,'' in {\em Proceedings of the Seventh Annual
  ACM/IEEE International Conference on Mobile Computing and Networking
  (Mobicom)}, 2001.

\bibitem{fangLocating2006}
Q.~Fang, J.~Gao, and L.~J. Guibas, ``Locating and bypassing holes in sensor
  networks,'' {\em Mob. Netw. Appl.}, vol.~11, no.~2, pp.~187--200, 2006.

\bibitem{fekete06}
A.~Kr\"oller, S.~P. Fekete, D.~Pfisterer, and S.~Fischer, ``Deterministic
  boundary recognition and topology extraction for large sensor networks,'' in
  {\em SODA '06: Proceedings of the seventeenth annual ACM-SIAM symposium on
  Discrete algorithm}, (New York, NY, USA), pp.~1000--1009, ACM Press, 2006.

\bibitem{wangBoundaryMobiCom06}
Y.~Wang, J.~Gao, and J.~S. Mitchell, ``Boundary recognition in sensor networks
  by topological methods,'' in {\em MobiCom '06: Proceedings of the 12th annual
  international conference on Mobile computing and networking}, (New York, NY,
  USA), pp.~122--133, ACM Press, 2006.

\bibitem{zv05}
A.~Zemlianov and G.~de~Veciana, ``Capacity of ad hoc wireless networks with
  infrastructure support,'' {\em IEEE Journal on selected areas in
  Communications}, vol.~23, March 2005.

\bibitem{HWKZL05}
X.~Hong, P.~Wang, J.~Kong, Q.~Zheng, and J.~Liu, ``Effective probabilistic
  approach protecting sensor traffic,'' {\em Military Communications
  Conference, 2005. MILCOM 2005. IEEE}, vol.~1, pp.~169--175, October 2005.

\bibitem{ward97new}
A.~Ward, A.~Jones, and A.~Hopper, ``A new location technique for the active
  office,'' 1997.

\bibitem{priyantha00cricket}
N.~Priyantha, A.~Chakraborty, and H.~Balakrishnan, ``The cricket
  location-support system,'' in {\em Proceedings of the 6th Annual ACM
  International Conference on Mobile Computing and Networking (MobiCom '00)},
  August 2000.

\bibitem{bhardwajChandraksanInfocom02}
M.~Bhardwaj and A.~Chandrakasan, ``Bounding the lifetime of sensor networks via
  optimal role assignments,'' in {\em Proceedings of the Twenty-First Annual
  Joint Conference of the IEEE Computer and Communications Societies (INFOCOM
  2002)}, vol.~3, pp.~1587--1596, 2002.

\bibitem{bloughSantiMobicom02}
D.~Blough and P.~Santi, ``Investigating upper bounds on network lifetime
  extension for cell-based energy conservation techniques in adhoc networks,''
  in {\em Proceedings of the Eighth Annual International Conference on Mobile
  Computing and Networking (ACM MobiCom 2002)}, 2002.

\bibitem{zhangHouMOBIHOC04}
H.~Zhang and J.~Hou, ``On deriving the upper bound of $\alpha$---lifetime for
  large sensor networks,'' in {\em MobiHoc '04: Proceedings of the 5th ACM
  international symposium on Mobile ad hoc networking and computing}, (New
  York, NY, USA), pp.~121--132, ACM Press, 2004.

\end{thebibliography}

\end{document}